\documentclass[a4paper]{scrartcl}
\pdfoutput=1
\usepackage[utf8]{inputenc}

\usepackage[T1]{fontenc} 
\usepackage{lmodern} 
\usepackage[sc,osf]{mathpazo} 
\AtBeginDocument{%
  \DeclareFontShape{T1}{pplj}{m}{scit}{<-> ec-qplri-sc}{}
}
\usepackage{myT1classico} 
\DeclareMathAlphabet{\mathsf}{T1}
  {\sfdefault}{m}{n} 
\SetMathAlphabet{\mathsf}{bold}{T1}{\sfdefault}{b}{n} 

\usepackage{amsmath,amsthm,mathtools}

\usepackage[british]{babel}
\usepackage{csquotes} 

\usepackage{microtype} 

\usepackage{authblk}

\usepackage[colorlinks]{hyperref}
\usepackage[nameinlink]{cleveref}

\usepackage[style=phys,biblabel=brackets,eprint=true,sorting=noney,
  sortcites=false]{biblatex}
\newbibmacro{string+doiurl}[1]{%
  \iffieldundef{doi}{%
    \iffieldundef{url}{#1}{\href{\thefield{url}}{#1}}%
  }{\href{https://doi.org/\thefield{doi}}{#1}}%
}
\DeclareFieldFormat{title}{\usebibmacro{string+doiurl}{\mkbibemph{#1}}}
\DeclareSortingTemplate{noney}{
  \sort{\citeorder}
}

\DefineBibliographyExtras{british}{}

\newcommand*{\D}{\mathrm{d}} 
\newcommand*{\R}{\mathbb R}

\theoremstyle{plain}
\newtheorem{theorem}{Theorem}

\makeatletter
\mathchardef\ordinarycolon\mathcode`\:%
\mathcode`\:=\string"8000%
\begingroup \catcode`\:=\active%
  \gdef:{\@ifnextchar{=}
    {\coloneq\@gobble}
    {\ordinarycolon}}%
\endgroup
\AtBeginDocument{\mathcode`\==\string"8000} 
\mathchardef\ordinaryequals\mathcode`\=%
\begingroup \catcode`\==\active%
  \gdef={\@ifnextchar{:}
    {\eqcolon\@gobble}
    {\ordinaryequals}
  }%
\endgroup
\makeatother

\title{\texorpdfstring{\vspace*{-1.5\baselineskip}}{}%
  The trace-free Einstein tensor is not variational for the metric as
  field variable}

\author[1,2,3,a]{Arian L. von Blanckenburg}
\author[3,4,b]{Domenico Giulini}
\author[3,c]{Philip~K.~Schwartz\footnote{Author to whom any
    correspondence should be addressed.}}
\affil[1]{Max Planck Institute for Gravitational Physics (Albert
  Einstein Institute), \par
  Callinstraße 38, 30167 Hannover, Germany}
\affil[2]{Leibniz University Hannover,
  Institute of Gravitational Physics, \par
  Callinstraße 38, 30167 Hannover, Germany}
\affil[3]{Leibniz University Hannover,
  Institute of Theoretical Physics, \par
  Appelstraße 2, 30167 Hannover, Germany}
\affil[4]{University of Bremen,
  Center of Applied Space Technology and Microgravity, \par
  Am Fallturm 1, 28359 Bremen, Germany}
\affil[a]{\normalfont\texttt{\href{mailto:arian.von.blanckenburg@aei.mpg.de}
    {arian.von.blanckenburg@aei.mpg.de}}}
\affil[b]{\normalfont\texttt{\href{mailto:giulini@itp.uni-hannover.de}
    {giulini@itp.uni-hannover.de}}}
\affil[c]{\normalfont\texttt{\href{mailto:philip.schwartz@itp.uni-hannover.de}
    {philip.schwartz@itp.uni-hannover.de}}}

\date{}

\begin{document}
\maketitle

\vspace{-3.5\baselineskip}
\enlargethispage{1.3\baselineskip}

\begin{abstract}
  \noindent
  It is well-known that the trace-free Einstein tensor of a
  pseudo-Riemannian metric cannot arise by variation of a local
  diffeomorphism-invariant action functional with the (inverse) metric
  as field variable.  We show that this statement remains true even
  for general local actions, without the assumption of diffeomorphism
  invariance.
\end{abstract}

The trace-free Einstein equations for a pseudo-Riemannian metric
$g_{\mu\nu}$ read
\begin{equation}
  \label{eq:tf_Einstein_eq}
  R_{\mu\nu} - \frac{1}{n} R g_{\mu\nu}
  = \kappa \left(T_{\mu\nu} - \frac{1}{n} T g_{\mu\nu}\right) \; ,
\end{equation}
where $n \ge 3$ is the number of spacetime dimensions, $R_{\mu\nu}$ is
the Ricci tensor of $g_{\mu\nu}$, $R$ is its Ricci scalar, $\kappa$ is
the Einsteinian gravitational constant (generalised to $n$
dimensions), $T_{\mu\nu}$ is the energy--momentum tensor of matter,
and $T$ its trace.  These equations are often considered an
interesting modification of the Einstein equations of general
relativity, for example due to the fact that when assuming matter
energy--momentum conservation $\nabla^\mu T_{\mu\nu} = 0$, they imply
the standard Einstein equations with a cosmological constant $\Lambda$
arising as an integration constant~\cite{Weinberg:1989,
  Ellis.EtAl:2011}. Conversely the standard Einstein equations imply
their trace-free part \eqref{eq:tf_Einstein_eq}.  However, when going
`back and forth' between the full and trace-free Einstein equations,
the value of $\Lambda$ can change: in the standard Einstein equations,
it is a fixed parameter, while in the trace-free case, arising as an
integration constant, it can have different values for different
solutions.

In this note, we are concerned with the trace-free Einstein tensor,
i.e.\ the left-hand side of \eqref{eq:tf_Einstein_eq}, which we denote
by
\begin{equation}
  G^\mathrm{tf}_{\mu\nu} := R_{\mu\nu} - \frac{1}{n} R g_{\mu\nu} \; .
\end{equation}
A well-known argument shows that $G^\mathrm{tf}_{\mu\nu}$---or, more
precisely, its densitised version, see below---cannot arise by
variation of a diffeomorphism-invariant local action functional with
respect to the inverse metric $g^{\mu\nu}$: due to diffeomorphism
invariance, such a variation has identically vanishing covariant
divergence (see, e.g., ref.~\cite[sec.\ 23]{Pauli:1921a,
  Pauli:1921b}\footnote{In ref.~\cite[sec.\ 23]{Pauli:1921a,
    Pauli:1921b} by Pauli, the only action functional treated
  \emph{explicitly} is the standard Einstein--Hilbert action of
  general relativity.  However, apart from the very last, concrete
  part of the computation, Pauli's analysis is completely general and
  shows, among other results, vanishing covariant divergence of the
  variation for general diffeomorphism-invariant local action
  functionals (even though this is not explicitly stated).}); however,
the trace-free Einstein tensor is \emph{not} identically
divergence-free (for $n \ge 3$), since the Einstein tensor $G_{\mu\nu}
= R_{\mu\nu} - \frac{1}{2} R g_{\mu\nu}$ is divergence-free by the
contracted second Bianchi identity
.

In the following, we are going to show that the same conclusion holds
true without the assumption of diffeomorphism invariance:
\begin{theorem}
  \label{thm:non-var}
  For $n \ge 3$, the densitised trace-free Einstein tensor
  \begin{equation}
    \label{eq:tf_Einstein_dens}
    \tilde{G}^\mathrm{tf}_{\mu\nu}
    := \sqrt{|\det(g_{\kappa\lambda})|} \, G^\mathrm{tf}_{\mu\nu}
    = \sqrt{|\det(g_{\kappa\lambda})|}
      \left(R_{\mu\nu} - \frac{1}{n} R g_{\mu\nu}\right)
  \end{equation}
  cannot arise by variation of \emph{any} local action functional with
  respect to the inverse metric.
\end{theorem}
\noindent
We remind the reader that a \emph{local functional} means an integral
over a Lagrangian density depending pointwise on the field variable
(i.e.\ in the present case the inverse metric) and finitely many of
its derivatives.

\enlargethispage{2.1\baselineskip}

Here we have specifically considered the \emph{densitised} version of
the trace-free Einstein tensor (as a tensor density of weight $1$)
since variation of any action functional with respect to a tensor
field (i.e.\ a tensor density of weight $0$) yields a tensor density
(of weight $1$): only the densitised tensor has the chance of arising
as a variational derivative.

Of course, the non-variationality of the trace-free Einstein tensor
need not be any obstacle towards consideration of the trace-free
Einstein equations \eqref{eq:tf_Einstein_eq} as a viable alternative
to the standard Einstein equations---for example, the discussion in
ref.\ \cite{Ellis.EtAl:2011} expressly only works with the equations,
not discussing variational formulations.

Note that for the interpretation of the non-variationality conclusion
of \cref{thm:non-var} it is crucial that the field variable considered
is the inverse metric alone, and that the result makes a statement on
the trace-free Einstein \emph{tensor}: the only conclusion which
\cref{thm:non-var} allows for the trace-free Einstein \emph{equations}
\eqref{eq:tf_Einstein_eq} is that the latter may not arise directly as
the Euler--Lagrange equations of some local functional with respect to
the inverse metric.  In fact, by reformulating the equations in terms
of different field variables and/or adding auxiliary dynamical fields,
several action principles for the trace-free Einstein equations have
been constructed recently \cite{Montesinos.Gonzalez:2023,
  Montesinos.Gonzalez:2024,Montesinos.Gonzalez:2025}---these actions
are even diffeomorphism invariant.

Related to this remark is the subject of unimodular gravity, in which
the trace-free Einstein equations are complemented by a condition
fixing the metric volume density $\sqrt{|\det(g_{\mu\nu})|}$.
Unimodular gravity is commonly discussed in terms of variational
principles; see, e.g., ref.\ \cite{Bengochea.EtAl:2023} for a
diffeomorphism-invariant discussion employing a fixed background
volume (pseudo-)form.  As above, \cref{thm:non-var} implies that the
trace-free Einstein equations cannot arise directly as the
Euler--Lagrange equations of any such (local) action functional
employed in unimodular gravity with respect to $g^{\mu\nu}$.  And
indeed, this is satisfied in variational formulations of unimodular
gravity, be it the volume-form formulation of ref.\
\cite{Bengochea.EtAl:2023} or more traditional formulations
constraining $\sqrt{|\det(g_{\mu\nu})|} = 1$ in a fixed coordinate
chart: variation with respect to the inverse metric yields directly
the \emph{full} Einstein equations, in which the role of the
cosmological constant is taken by the Lagrange multiplier implementing
the volume constraint (which may take an arbitrary value).  As
remarked at the beginning of the present paper, this is then
equivalent to the trace-free Einstein equations; these, however, do
\emph{not} directly arise by variation.

\enlargethispage{1.5\baselineskip}

We are going to prove \cref{thm:non-var} by methods from the inverse
calculus of variations, namely the so-called method of
\emph{(canonical) variational completion} \cite{Voicu.Krupka:2015,
  Hohmann.EtAl:2021}.  Informally formulated, this method works as
follows.  One considers a tensor-density-valued expression $E_A[y^B]$
depending pointwise on a tensor field $y^A$ and finitely many of its
derivatives, such that $E_A$ could in principle arise from variation
of some action functional with respect to $y^A$---i.e.\ the expression
$E_A$ takes values, apart from the densitising, in the dual of the
bundle in which $y^A$ lives, as indicated by our choice of indices.
To such an expression $E_A[y^B]$, there is an associated so-called
\emph{(extended) Vainberg--Tonti Lagrangian}, a scalar density defined
by a specific one-dimensional integral: let $a \in \R \cup
\{\pm\infty\}$ such that $\lim_{u \to a} u E_A[u y^B] = 0$.  The
Vainberg--Tonti Lagrangian associated to $E_A[y^B]$ is then
\begin{equation}
  \mathcal L[y^A] := y^B \int_a^1 E_B[u y^A] \D u \; .
\end{equation}
If no such $a$ exists or the integral does not converge, the
Vainberg--Tonti Lagrangian associated to $E_A[y^B]$ does not exist.
The fundamental result of variational completion now is the following
(see \cite[Theorem 1]{Hohmann.EtAl:2021}\footnote{In the original
  formulation of variational completion in ref.\
  \cite{Voicu.Krupka:2015}, the lower limit of the integral defining
  the Vainberg--Tonti Lagrangian was assumed to be $0$.  In ref.\
  \cite{Hohmann.EtAl:2021}, this was generalised to allow for a lower
  limit $a$ differing from $0$.  For notational simplicity, ref.\
  \cite{Hohmann.EtAl:2021} treated only the case of second-order
  expressions $E_A[y^B]$ (i.e.\ depending on up to second derivatives
  of $y^A$); however the proof of \cite[Theorem 1]{Hohmann.EtAl:2021}
  works analogously for arbitrary finite order.  In any case, for our
  present application second order is sufficient, since the trace-free
  Einstein tensor is of second order in the inverse metric.}):
\begin{theorem}[informal statement]
  If the Vainberg--Tonti Lagrangian $\mathcal L[y^A]$ associated to
  $E_A[y^B]$ exists and the expression $E_A[y^B]$ is locally
  variational, then variation of the action $\int \mathcal L[y^A] \D^n
  x$ with respect to $y^A$ will yield the original expression
  $E_A[y^B]$.  \qed
\end{theorem}
Put differently: for $E_A[y^B]$ locally variational, i.e.\ arising
from some locally defined Lagrangian depending pointwise on $y^B$ and
its derivatives, the Vainberg--Tonti Lagrangi\-an---if it exists---is
a Lagrangian equivalent to the original one.  Taking the
contrapositive, we obtain the following:
\begin{theorem}[informal statement]
  \label{thm:var_compl_contrap}
  If the Vainberg--Tonti Lagrangian $\mathcal L[y^A]$ associated to
  $E_A[y^B]$ exists and variation of the action $\int \mathcal L[y^A]
  \D^n x$ with respect to $y^A$ yields an expression \emph{different}
  from $E_A[y^B]$, then $E_A[y^B]$ cannot be locally variational.
  \qed
\end{theorem}

By applying \cref{thm:var_compl_contrap}, we can now easily prove
non-variationality of the trace-free Einstein tensor:

\begin{proof}[Proof of \cref{thm:non-var}]
  We consider the densitised trace-free Einstein tensor
  \eqref{eq:tf_Einstein_dens} as an expression depending on the
  inverse metric and its derivatives, i.e.\
  $\tilde{G}^\mathrm{tf}_{\mu\nu}[g^{\rho\sigma}]$.  In order to
  compute the Vainberg--Tonti Lagrangian associated to
  $\tilde{G}^\mathrm{tf}_{\mu\nu}[g^{\rho\sigma}]$ , we have to
  determine how it behaves when the inverse metric is scaled with $u
  \in \R \setminus \{0\}$.  (Here we have to exclude the value $u = 0$
  in order to stay in the set of pseudo-Riemannian metrics.)

  By definition, when the inverse metric is scaled with $u$, the
  metric is scaled with $u^{-1}$.  Thus the Christoffel symbols are
  invariant under this scaling, likewise the Ricci tensor.  Hence the
  Ricci scalar $R = g^{\mu\nu} R_{\mu\nu}$ is scaled with $u$.
  Finally, $\sqrt{|\det(g_{\kappa\lambda})|}$ is scaled with
  $|u|^{-n/2}$.  Combined, we obtain
  \begin{equation}
    \tilde{G}^\mathrm{tf}_{\mu\nu}[u g^{\rho\sigma}]
    = |u|^{-n/2} \sqrt{|\det(g_{\kappa\lambda})|}
      \left(R_{\mu\nu} - \frac{1}{n} u R u^{-1} g_{\mu\nu}\right)
    = |u|^{-n/2} \tilde{G}^\mathrm{tf}_{\mu\nu}[g^{\rho\sigma}] \; .
  \end{equation}
  Thus, there is a unique $a \in \R \cup \{\infty\}$ satisfying
  $\lim_{u \to a} u \tilde{G}^\mathrm{tf}_{\mu\nu}[u g^{\rho\sigma}] =
  0$, namely $a = \infty$: for $n \ge 3$, the overall scaling factor
  $u |u|^{-n/2} = \mathrm{sgn}(u) |u|^{-(n-2)/2}$ vanishes only as $u
  \to \infty$.

  Hence, the Vainberg--Tonti Lagrangian associated to the densitised
  trace-free Einstein tensor as a functional of the inverse metric is
  \begin{align}
    \mathcal L[g^{\mu\nu}]
    &= g^{\rho\sigma} \int_\infty^1
      \tilde{G}^\mathrm{tf}_{\rho\sigma}[u g^{\mu\nu}] \D u \nonumber\\
    &= g^{\rho\sigma} \tilde{G}^\mathrm{tf}_{\rho\sigma}[g^{\mu\nu}]
      \int_\infty^1 u^{-n/2} \D u \nonumber\\
    &= -\frac{2}{n-2} \,
      g^{\rho\sigma} \tilde{G}^\mathrm{tf}_{\rho\sigma}[g^{\mu\nu}]
      \nonumber\\
    &= 0 \; .
  \end{align}
  In the third step, we have used that $n \ge 3$ in the computation of
  the integral, and in the last step that
  $\tilde{G}^\mathrm{tf}_{\mu\nu}$ is trace-free, i.e.\ $g^{\mu\nu}
  \tilde{G}^\mathrm{tf}_{\mu\nu} = 0$.

  Since the Vainberg--Tonti Lagrangian vanishes, variation of the
  corresponding action with respect to $g^{\mu\nu}$ yields $0$.
  Hence, according to \cref{thm:var_compl_contrap}, the densitised
  trace-free Einstein tensor is not variational for the inverse metric
  as field variable.
\end{proof}

Note that the result stays true when considering the
\emph{contravariant} trace-free Einstein tensor $R^{\mu\nu} -
\frac{1}{n} R g^{\mu\nu}$ for the \emph{metric} $g_{\mu\nu}$ as field
variable: The scaling changes to the metric being scaled with $u$ and
the inverse metric with $u^{-1}$, leading again to invariant
Christoffel symbols and Ricci tensor, such that the Ricci scalar is
scaled with $u^{-1}$ and the determinant factor with $|u|^{n/2}$.
Combined, the densitised contravariant trace-free Einstein tensor is
scaled with $|u|^{n/2}$, which yields $a = 0$ as the lower integration
limit in the integral in the definition of the Vainberg--Tonti
Lagrangian.  In the end, due to the trace-freeness, the
Vainberg--Tonti Lagrangian again vanishes.

\printbibliography

\end{document}
